\documentstyle[12pt]{article}

\date{}
\begin{document}
\title{Dilatons in the Topological Soliton Model for the Hyperons}

\author{K. Tsushima and D.O. Riska}
\maketitle
\centerline{\it Research Institute for Theoretical Physics and}

\centerline{\it Department of Physics, University of Helsinki,
SF-00170 Helsinki, Finland}

\setcounter{page}{0}
\vspace{1cm}

\centerline {\bf Abstract}

We show that the predicted hyperon masses in the topological soliton
model are very sensitive to the value of the gluon condensate parameter
that appears when the scale invariance and trace anomaly of QCD are
taken into account by introduction of a dilaton field. This contrasts with
the insensitivity of the soliton properties to the dilaton coupling.
In order that the predicted strange and charmed hyperon spectra agree
with the empirical ones the gluon condensate parameter has to be about
(400 MeV)$^4$, which agrees with the result obtained from QCD sum rules. This
implies that the bag formed by the scalar field must be very shallow.
\newpage

{\bf 1. Introduction}\\

The extension to the heavy flavour sectors of the Skyrme model [1],
in which the hyperons are described as bound states of heavy flavour
isodoublet mesons and a soliton formed of the light (u, d) flavours
[2, 3], leads to remarkably good predictions of the hyperon properties
[4, 5]. The spectra and magnetic moments of the strange, charm and
bottom hyperons are close to the available empirical values, and also
to the corresponding quark model based predictions [5, 6]. This bound
state model, which incorporates chiral symmetry for the light flavour
(SU(2)) sector, also to a good approximation respects the constraints
imposed by heavy quark symmetry in the baryon sector [7], although in
its original form it does not include the heavy flavour vector meson
fields.\\

The topological soliton model can also be modified so as to
incorporate the scale invariance and trace anomaly of the underlying
QCD, for which it is intended to describe in the large
colour number approximation. The
simplest way to achieve this is to introduce a scalar or dilaton
field, which should model the vacuum fluctuations of the gluon field
[8]. This leads to the formation of a "bag" around the soliton, where
the dilaton field is nonzero. The depth of this scalar bag will
depend on the model used for the soliton Lagrangian. When the original
Skyrme model Lagrangian is used it has been found that the sensitivity
of the predicted
nucleon properties to the dilaton coupling, and the two parameters
that govern its strength - the gluon condensate parameter and the
scalar meson (glueball) mass - is rather weak [8].\\

We here study the modification of the bound state model for the
hyperons caused by the coupling to the dilaton field. As the dilaton
field leads to a modification of both the soliton profile and the wave
equation for the heavy flavour mesons, the predicted values for the
hyperon masses are far more sensitive to the presence of the
dilaton field than the nucleon properties, which only depend on the
SU(2) soliton. We find that the main effect of the coupling to the
dilaton field is to increase the binding energy of the heavy flavour
mesons, and thus to lower the hyperon masses. This is mainly due to
the reduction of the mass term in the meson wave equation at short
range. Retention of the successful predictions of the hyperon
phenomenology achieved in ref. [4-6] therefore requires that the
coupling to the dilaton field be weak and that the bag formed by the
scalar field be very shallow. The corresponding value for the
gluon condensate has to be of the order (400 MeV)$^4$, in agreement
with QCD sum rules. The results are less sensitive to the glueball
mass, but the
nonobservation of any low lying glueball suggests that it has to be
at least 1.5 GeV.\\

In section 2 of this paper we show how the bound state model for the
hyperons is modified by the coupling to the dilaton field. In section
3 we show the numerical results for the soliton profile and the scalar
field amplitude obtained with different values for the
gluon condensate. In section 4 we show how the introduction of the
dilaton field affects the predicted hyperon properties, and
demonstrate the sensitivity to the
gluon condensate value. Section 5 contains a concluding
discussion.\\

{\bf 2. Dilaton coupling in the bound state model}\\

The bound state model for the hyperons is based on the Skyrme model [4,
5] or some extension thereof, to include the heavy flavours [10]. The
Lagrangian density of this model does not have the scale invariance
of the classical QCD Lagrangian density. The required scale invariance
can be introduced by coupling of a scalar dilaton field to those terms
in the Lagrangian density that break the scale invariance. The
divergence of the classically conserved Noether current of the scale
transformation is finally determined by the trace anomaly of QCD at
the quantum level.
[8,9].\\

We here consider the original Skyrme model extended to include the
heavy flavour sectors. Denoting the soliton field U and the scalar
dilaton field $\sigma$, the scale invariant basic Lagrangian density
is
$${\cal L}= e^{2\sigma} \{{1\over2} \Gamma_0^2 \partial_\mu \sigma
\partial^ \mu \sigma + {1\over 4} f_\pi^2 Tr[\partial_\mu U \partial ^
\mu U ^ \dagger]\}$$
$$+{1\over 32 e^2 } Tr [L_\mu, L_\nu]^2 + {\cal L}_{CSB} - V (\sigma),
\eqno (2.1)$$\\
\noindent
with $L_\mu=U^\dagger \partial_\mu U$. Here ${\cal L}_{CSB}$
is the charge symmetry breaking mass term, which for the case that $U$
is an SU(3) field takes the form [11]

$${\cal L}_{CBS}={f_\pi^2 m_\pi^2+2f^2 m^2 \over 12} e^{3\sigma} Tr \{U+U^
\dagger -2\} + {f_\pi^2 m_\pi^2 - f^2 m^2 \over 6}e^{3\sigma} Tr\{\sqrt
3 \lambda_8(U+U^ \dagger)\}$$
$$-{f_\pi^2-f^2 \over 12}e^{2\sigma} Tr \{(1-\sqrt 3 \lambda_8)(U \partial
_\mu U^ \dagger \partial ^\mu U + U^ \dagger \partial _\mu U \partial
^\mu U^ \dagger)\}. \eqno (2.2)$$\\
\noindent
Here the $m_\pi$ denotes the mass of the pion and $m$ that of the heavy
flavour meson and $f_\pi$ and $f$ the corresponding decay constants.
The potential function $V(\sigma)$ in (2.1) is defined as

$$V(\sigma)={1\over 4} C_G e^{4\sigma}(4 \sigma-1), \eqno (2.3)$$\\
\noindent
where $C_G$ is the value of the gluon condensate. Finally the
parameter $\Gamma_0$ in (2.1) is determined by the value of the gluon
condensate $C_G$ and the glueball mass $m_G$ as

$$\Gamma_0={2 \over m_G} \sqrt {C_G}. \eqno (2.4)$$\\
\noindent
The factor $e^{2\sigma}$ in the bilinear derivative terms in the
Lagrangian density introduces the scale invariance. The factor
$e^{3\sigma}$ in the mass term is introduced for similarity with the
quark mass term contribution in the trace anomaly equation if the
anomalous dimension is taken to be zero [9]. As
will be shown below, since $\sigma<0$, this factor that reduces the mass
term contribution at short range has a strong lowering effect on the
predicted hyperon masses if $\sigma$ deviates appreciably from 0.\\
\noindent
The Lagrangian (2.1) has finally to be completed with the
Wess-Zumino action

$$S=-i{N_C \over 240 \pi^2} \int d^5 x \epsilon^{\mu \nu \alpha \beta
\gamma} Tr[L_\mu L_\nu L_\alpha L_\beta L_\gamma], \eqno (2.5)$$\\
\noindent
which is scale invariant by itself. This term does not contribute
to the energy of the system at the level of SU(2) but it does lead to
an important contribution in the meson wave equation.\\
\noindent
For the SU(3) field $U$ we adopt the form [12]

$$U= \sqrt {U_M} U_\pi \sqrt {U_M}. \eqno (2.6)$$\\
\noindent
Here $U_\pi$ is the soliton field

$$U_\pi= \left (\begin{array} {c c} u & 0\\ 0 & 1 \end{array} \right)
\eqno (2.7)$$\\
\noindent
with $u$ being the SU(2) hedgehog field that describes the Skyrmion:

$$u=e^{i\vec\tau \cdot \hat r \theta(r)}, \eqno (2.8)$$\\
\noindent
The chiral angle $\theta(r)$ is determined by the Euler-Lagrange
equation that corresponds to the Lagrangian density (2.1), (2.2). This
coupled differential equation for the functions $\theta(r)$ and
$\sigma(r)$ is given below.\\
\noindent
The heavy flavour meson field $U_M$ has the form

$$ U_M=exp \{ \frac{i \sqrt 2}{f} \left (\begin {array}{cc} 0 & M \\ M^
\dagger & 0 \end{array} \right)\}. \eqno (2.9)$$\\
\noindent
Here $M$ is one of the $S=-1$, $C=+1$ or $B=-1$ doublets

$$M= \left(\begin {array} {c} K^+\\ K^0  \end{array} \right),
\left( \begin{array}{c} \bar D^0 \\ D^- \end{array} \right),
\left(\begin{array}{c} B^+ \\ B^0 \end{array} \right). \eqno (2.10)$$\\
\noindent
When the field expression (2.6) is inserted into the Lagrangian
density (2.1), (2.2) (and the Wess-Zumino action (2.5)), and this is
expanded to second order in the heavy flavour field $U_M$, a linear wave
equation for the meson field modes is obtained. Upon a rescaling of the
meson field this wave equation takes the form [12]

\newpage
$$a(r) \nabla^2 M+b(r) \hat r \cdot \vec\nabla M - c(r) \vec L^2 M$$
$$-[v_0(r)+v_{IL}(r) \vec I \cdot \vec L] M - e^{3\sigma} m^2 M +d(r)
\omega ^2 M +2 \omega \lambda(r) M=0. \eqno (2.11)$$\\
\noindent
Here $\vec I$ is the effective meson spin operator, $\vec L$ the
orbital angular
momentum operator and $\omega$ the meson energy. The radial functions
$a(r)$...$\lambda(r)$ are

$$a(r)=e^{2\sigma}+\frac{1}{2e^2f^2} \frac {sin^2 \theta}{r^2}, \quad
b(r)=\frac {1}{2e^2 f^2 r} [\frac {sin2 \theta}{r} \theta' -2 \frac
{sin^2\theta} {r^2}] + 2 e^{2\sigma} \sigma',$$
$$c(r)=\frac {1}{4e^2 f^2 r^2} (\theta^{'2} - \frac{sin^2 \theta}{r^2}), \quad
d(r)=e^{2 \sigma} + \frac {1}{4e^2 f^2} (\theta^{'2}+2 \frac
{sin^2 \theta}{r^2}),$$
$$\lambda(r)=-\frac{3}{8\pi^2 f^2} \frac {sin^2 \theta}{r^2} \theta'.
\eqno (2.12)$$\\
\noindent
The potential functions $v_0(r)$ and $v_{IL}(r)$ are

$$v_0(r)=-\frac {1}{2} (\theta^{''} tan \frac {\theta}{2}+\frac
{\theta^{'2}}{2}) (e^{2\sigma}+\frac{1}{2e^2 f^2} \frac {sin^2
\theta}{r^2})$$
$$ - \frac {\theta'}{r} tan \frac {\theta}{2}\: [(1+r
\sigma') e^{2\sigma} + \frac {1}{4e^2 f^2} \frac {\theta' sin2
\theta}{r}]$$
$$\frac {1}{2}(1-\frac{f_\pi^2}{f^2}) e^{2\sigma} tan \frac{\theta}{2}\:
[2\sigma' \theta' + \theta^{''} +\frac{2\theta'}{r} -\frac
{sin2\theta}{r^2}]. \eqno (2.13a)$$\\
\noindent
$$v_{IL}(r)=\frac{4sin^2 \frac{\theta}{2}} {r^2} [e^{2\sigma} +
\frac{1}{e^2 f^2} (\theta^{'2} + \frac{sin^2\theta}{r^2})]$$
$$-\frac{3}{2e^2 f^2 r^2} [\frac{sin^2 \theta}{r^2} -
\theta^{'2}(1-4sin^2\frac {\theta}{2}) -\theta^{''}sin\theta]. \eqno
(2.13b)$$\\
\noindent
These functions are the same as those derived in ref. [12], except for
the additional factors that involve the scalar field $\sigma$ in the
terms in the radial functions that arise from the quadratic term
in the Lagrangian density (2.1) and the factor
$e^{3\sigma}$ in the mass term.\\

The hyperon mass is obtained as the sum of the mass of the SU(2) soliton,
the meson energy and hyperfine structure correction. The expression
for the mass of the soliton coupled to the dilaton field is

$$M=\pi \int dr [2 \Gamma_0^2 e^{2 \sigma} r^2 \sigma^{'2}+2f_\pi^2
e^{2\sigma} (r^2
\theta ^{'2} + 2sin^2 \theta) + \frac {2}{e^2} sin^2
\theta (2\theta ^{'2} + \frac {1}{r^2} sin^2\theta)$$
$$+4f_\pi^2 m_\pi^2 e^{3\sigma} r^2 (1-cos \theta) + C_G r^2(e^{4\sigma}
(4\sigma -1) +1)]. \eqno (2.14)$$\\
\noindent
The general expression for the hyperfine structure correction to the
hyperon mass is given in ref. [5]; the only modification being the
insertion of the factor $e^{2\sigma}$ in all the terms in the expression for
the hyperfine structure constant, which do not involve the factor
$e^{-2}$ - i.e. the terms that arise from the stabilizing term in the
Lagrangian density (2.1).\\

By requiring the soliton mass (2.14) to be stationary one obtains the
coupled equations of motion for the functions $\theta$ and $\sigma$ as:

$$ f_\pi^2 e^{2\sigma}(sin2 \theta - 2 r^2 \sigma' \theta' -2r \theta' -
r^2 \theta^{''}) + \frac {1}{e^2}(\frac{1}{r^2} sin^2 \theta sin2
\theta - \theta^{'2} sin2\theta- 2 \theta^{''} sin^2 \theta)$$
$$+f_\pi^2 m_\pi^2 e^{3\sigma} r^2 sin \theta =0, \eqno (2.15a)$$\\

$$f_\pi^2 (r^2 \theta^{'2} + 2 sin^2 \theta) - \Gamma_0^2 (2r \sigma'
+ r^2 \sigma^{'2} + r^2 \sigma^{''}) + 3f_\pi^2 m_\pi^2 e^\sigma r^2
(1-cos \theta)$$
$$+4C_G e^{2\sigma} r^2 \sigma=0. \eqno (2.15b)$$\\
\noindent
The only parameters in the model are the pion and heavy meson decay
constants $f_{\pi}$ and $f$, the inverse strength of the stabilizing
term in the soliton Lagrangian $e$ and the value for the gluon
condensate $C_G$ and the glueball mass $m_G$.\\

{\bf 3. The soliton parameters}\\

In order that meaningful predictions for the hyperon spectra be
obtained the parameters in the Lagrangian density (2.1) should be
chosen so that the nucleon and the $\Delta_{33}$ resonance take their
empirical values. This leaves two of the parameters free. It is
natural to choose the value of the gluon condensate $C_G$  and the
glueball mass to agree with available results from QCD sum rules and
lattice gauge calculations and then to vary $f_\pi$ and $e$ until the
correct nucleon and $\Delta_{33}$ mass values are obtained. The
nonobservation of any low lying glueball suggests that the lowest
possible value for the mass of the glueball is at least 1.5 GeV. In
order that the dilaton coupling not be insignificantly small we shall
choose this value for $m_G$. We have verified numerically that
increasing this value to 2.0 GeV does not change the result significantly.\\

The average of the values for the gluon condensate parameter with the factor
9/8 obtained by QCD sum
rules is $C_G \simeq$ (392 MeV)$^4$ [13]. With these values for $m_G$ and
$C_G$ we find that the remaining parameters $f_\pi$ and $e$ should be
53.63 MeV and 4.795 respectively, in order that the nucleon and the
$\Delta_{33}$ resonance mass take their empirical values. These two
values are close to those for the case of no dilaton field ($f_\pi$=54
MeV, $e$=4.84) [14]. For these parameter values the scalar bag that is
formed by the dilaton field is very shallow. The functions $\theta(r)$
and $\sigma(r)$ obtained using these parameter values are shown in
Fig. 1.\\

Smaller values for the gluon condensate  parameter leads to a deeper bag
formation. The concomitant shrinkage of the soliton profile
$\theta(r)$ leads to poorer values for the nuclear radii and axial
coupling constant however [9], although the nucleon and $\Delta_{33}$
mass values may still have their empirical values. Results of lattice
QCD calculations suggest somewhat smaller values for the gluon
condensate ($C_G \simeq$ (291 MeV)$^4$) [15], than the QCD sum rules.
This still leads as a shallow bag. However, we shall
below show that much lower values for $C_G$ would lead to very poor
predictions for the hyperon masses. For example, choosing the small value
$C_G$=(180 MeV)$^4$, which leads to a deep bag, and $m_G$=1.5 GeV we
find that the nucleon and $\Delta_{33}$ masses agree with the
empirical ones if $f_\pi$=52.25 MeV and $e$=4.412. The corresponding
values for the chiral angle $\theta$ and scalar field $\sigma$ are
also plotted in Fig. 1.\\

{\bf 4. The influence of the scalar field on the hyperon masses}\\

The parameter choice that leads to the shallow scalar bag implies that
the coupling to the scalar meson field is very weak and that
consequently the soliton profile is very close to that obtained in the
case of no scalar field. As a consequence the solutions to the
wave equation (2.11) are also very close to those obtained in the case
$\sigma$=0. Quantitatively the relative change in the meson energy
caused by the weakly coupled $\sigma$-field is about twice that of
the corresponding maximal relative change in the chiral angle. This
proportionality however only holds when the absolute value of $\sigma$
is less than about 0.1.\\

In the case of the strange pseudoscalar isodoublet, with $f=f_K=1.23
f_\pi$, and using the parameter values that correspond to the shallow bag
case, the meson energy in the ground (P-) state is 205 MeV,
as compared to the value 209 MeV
that obtains when $\sigma=0$ [5]. In the case of the D-meson,
with $f_D=2$ $f_\pi$
the meson energy is 1306 MeV as compared to the value 1342 MeV
obtained in the absence of the dilaton. A similar small relative
shift of the B-meson energy from the value 3773 MeV given in ref. [5]
to 3665 MeV is caused by the shallow scalar bag. It is therefore
evident that a weak dilaton coupling of this type, which is
obtained with values for the gluon condensate and the glueball mass
obtained from QCD sum rules and phenomenological analyses, has no real
numerical
significance for the predicted hyperon masses.\\

A weak coupling of the dilaton field of this type has an almost
negligible effect on the hyperfine correction to the hyperon mass,
which is responsible for the isospin splitting of the hyperon masses.
For the ground state the calculated hyperfine splitting constant $c$
retains the value 0.39 that is obtained in the absence of the dilaton
field [5]. The predicted values for the masses of the $\Lambda$(1116) and
$\Sigma$(1193) hyperons thus in the presence of a weakly coupled
dilaton are 1082 MeV and 1201 MeV as compared to the values 1086 MeV
and 1297 MeV in the absence of a dilaton [5].\\

In the case of the parameter choices that correspond to a deep scalar
bag above the chiral angle remains close to the original values for
the case $\sigma=0$ (Fig. 1.). The large negative values for
$\sigma$ at short ranges do however have a disastrous influence
on the predicted
hyperon masses. The main reason for this is the reduction of the
strength of the meson mass term in the wave equation (2.11) that is
caused at short ranges by the factor $e^{3\sigma}$. In the case
of the kaons, with $f_K=1.23 f_\pi$
as above, the meson energy obtained by solving (2.11) using the
functions $\theta$ and $\sigma$ that correspond to the deep bag case,
drops to only 76 MeV from the original value 209 MeV
[5]. This would imply an underbinding of more than 100 MeV for the
kaons. The deep bag model also leads to a poor value for the hyperfine
splitting constant. In the ground state its value increases from 0.39
to 0.64.\\

The problem becomes even sharper in the case of the D-meson. With
$f_D=2 f_\pi$ (as in ref. [2]), the D-meson energy obtained by solving
the wave equation (2.11) for the deep bag case drops to 374 MeV. This
large underbinding of 1 GeV is clearly unrealistic and rules out the deep bag
parameters. The corresponding drop of the predicted B-meson energy
would be from the original value of 3773 MeV to only 730 MeV. The deep
bag model leads to too small values for the hyperfine structure
constants for the heavy flavour mesons. In the case of the B-meson
ground state it becomes negative.\\

{\bf 5. Discussion}\\

The results obtained above show that the hyperon masses predicted by
the bound state version of the topological soliton model are very
sensitive to the coupling to the dilaton field and in particular to
the value for the gluon condensate. In contrast the soliton profile -
and consequently also the predicted nucleon observables - are rather
weakly dependent on the strength of the dilaton coupling and thus do
not by themselves rule out a deep scalar bag at short range. The
sensitivity of the predicted hyperon masses (or heavy flavour meson
energies) to the parameters for the scalar field, and the very
unrealistic mass values obtained with a deep bag at short range do however
clearly rule out a deep bag model.\\

It is satisfying that the values for the gluon condensate and glueball
mass required for the shallow bag formation agree with those obtained
from QCD based analyses and particle phenomenology [9]. The bound
state hyperon model [2, 3] is thus robust in this sense. The gain in
the model that is achieved by introduction of the dilaton field in
this minimal way is
however mainly decorative in the theoretical sense, as it is
numerically insignificant although it is important in principle to
build in the scale invariance of QCD.\\

The hyperon mass values predicted in refs. [4,5] remain almost
unmodified by the presence of a weakly coupled dilaton field. The
predicted masses of the strange and charmed hyperons thus remain
in good agreement with
the available empirical values. In the case of the bottom hyperons the
predicted mass values are too low by $\simeq$ 600 MeV. The
reduction in binding energy required to overcome this problem can
probably only by achieved by explicit
coupling of the heavy flavour vector meson fields. Introduction of the
heavy vector meson fields is important in principle as well, as heavy
quark symmetry implies a degeneracy of the heavy pseudoscalar and
vector mesons and that they be treated in a similar way, with the same
coupling strength to baryonic matter.

\newpage
{\bf References}
\vspace{1cm}
\begin{enumerate}
\item T.H.R. Skyrme, Proc. Roy. Soc. {\bf A260} (1961) 127
\item C.G. Callan and I. Klebanov, Nucl. Phys. {\bf B202} (1985) 365
\item C.G. Callan, K. Hornbostel and I. Klebanov, Phys. Lett.
{\bf B202} (1988) 269
\item D.O. Riska and N.N. Scoccola, Phys. Lett {\bf 265} (1991) 198
\item M. Rho, D.O. Riska and N.N. Scoccola, Z. Phys. {\bf A341} (1992)
343
\item Y. Oh, D.P. Min, M. Rho and N.N. Scoccola, Nucl. Phys. {\bf
A534} (1991) 493
\item N. Isgur and M.B. Wise, Phys. Lett. {\bf B232} (1989) 113
\item H. Gomm, P. Jain, R. Johnson and J. Schechter, Phys. Rev. {\bf
D33} (1986) 3476
\item P. Jain, R. Johnson and J. Schechter, Phys. Rev. {\bf D35}
(1987) 2230
\item N.N. Scoccola, H. Nadeau, M. Nowak and M. Rho, Phys. Lett. {\bf
B201} (1988) 425
\item G. Pari, B. Schwesinger and H. Walliser, Phys. Lett. {\bf B255}
(1991) 1
\item U. Blom, K. Dannbom and D.O. Riska, Nucl. Phys. {\bf A493}
(1989) 384
\item M.A. Shifman, A.I. Vainshtein and V.I. Zakharov, Nucl. Phys.
{\bf B147} (1979) 385
\item G.S. Adkins and C.R. Nappi, Nucl. Phys. {\bf B233} (1984) 109
\item H. Satz, Phys. Lett. {\bf B113} (1982) 245
\end{enumerate}
\vspace{1cm}

\newpage
{\bf Figure Captions}\\

\noindent
Fig. 1.	The chiral angle $\theta$ and scalar field $\sigma$ as functions of
$r$ (in fm) for the case of a shallow (solid curve) and deep (dashed curve)
scalar bag.
\end{document}